\documentclass[a4paper,10pt]{article}

\usepackage[english]{babel}
\usepackage[T1]{fontenc}
\usepackage[dvips]{graphicx}
\usepackage{amsmath, amsfonts,amsthm, mathrsfs, amssymb}
\usepackage{verbatim}

\usepackage{enumerate}
\usepackage[latin1]{inputenc}
\usepackage{geometry}

\def\sidjd{(1+| i^d-j^d|)}

\def\sgeq{\succeq}

\def\ir{{\rm i}}

\def\diag{{\rm diag}}

\def\ops{OPS}

\def\norma#1{\left\|#1\right\|}
\def\sleq{\preceq}

\newcommand{\N}{{\mathbb N}}

\newcommand{\R}{{\mathbb R}}

\newcommand{\T}{{\mathbb T}}
\newcommand{\Z}{{\mathbb Z}}

\newcommand{\cB}{{\mathcal B}}

\newcommand{\cH}{{\mathcal H}}

\newcommand{\norm}[1]{\| #1 \|}

\newcommand{\im}{{\rm i}}

\def\uno{{\bf 1}}

\numberwithin{equation}{section}
\newtheorem{theorem}{Theorem}[section]
\newtheorem{lemma}[theorem]{Lemma}
\newtheorem{corollary}[theorem]{Corollary}

\newtheorem{definition}[theorem]{Definition}
\newtheorem{remark}[theorem]{Remark}

\def\Umag{{U}^{(1)}\null}

\def\utre{U^{(3)}\null}
\def\uquat{U^{(2)}\null}

\title{Reducibility  of 1-d Schr\"odinger equation with unbounded
  time quasiperiodic perturbations, III
}

\author{D. Bambusi\footnote{Dipartimento di Matematica, Universit\`a degli Studi di Milano, Via Saldini 50, I-20133
Milano. \newline
 \textit{Email: } \texttt{dario.bambusi@unimi.it}},
 R. Montalto\footnote{Institut f\"ur Mathematik, University of Z\"urich, Winterthurerstrasse 190, 8051, Z\"urich, Switzerland \newline 
 \textit{Email: } \texttt{riccardo.montalto@math.uzh.ch}}}

\begin{document}

\maketitle

\begin{abstract}
In this paper we study reducibility of time quasiperiodic
perturbations of the quantum harmonic or anharmonic oscillator in one space
dimension. We modify known algorithms obtaining a reducibility result
which allows to deal with perturbations of order strictly larger than the ones considered
in all the previous papers. 
\end{abstract}

\section{Introduction}

In this paper we study the reducibility of the time dependent
Schr\"odinger equation
\begin{align}
\label{schro}
\ir \dot\psi&=H(\omega t)\psi\ , 
\\
\label{H}
H(\omega t)&:=(-\im\partial_x-\epsilon W_1(x,\omega t))^2 + V(x)+\epsilon W_0(x,\omega
t)  \ ,\quad \ x\in\R\ ,
\end{align}
where $V$ is a smooth potential growing as $V(x)\simeq |x|^{2\ell}$, $\ell\geq1$, as $x\to\infty$, and
${W}_i$ are real valued functions (symbols) of class $C^\infty$,
depending in a quasiperiodic way on time. More precisely, we prove the
existence of a unitary (in $L^2$) transformation depending in
quasiperiodic way on time, which
conjugates the system to a diagonal time independent one. With
respect to previous results we allow here a more general class of
perturbations, including in particular the case of a harmonic
oscillator subject to a magnetic forcing, which was excluded by
previous papers.

From a physical point of view the main consequence is that a time
quasiperiodic perturbation of the kind considered here does not transfer
indefinitely energy to a quantum particle. From a mathematical point of view, this is expressed by the fact that the Sobolev norms of the solutions of \eqref{schro} stay bounded for all time. 
We recall that \eqref{H} was also studied for more
theoretical reasons: it is well known that the classical Duffing
oscillator, namely the Hamiltonian system with Hamiltonian
$\xi^2+x^4+\epsilon x^{\beta_0} \cos(\omega t)$, exhibits small
cahotic islands when $\epsilon\not=0$. The question is whether the
quantum system has some behaviors which are a quantum counterpart of
this nonregular behavior. Furthermore, a point of interest is whether
this depends on the value of the exponent $\beta_0$ or not. As a
consequence of reducibility, one gets that the quantum 
perturbed system qualitatively behaves forever as the unperturbed one, in
sharp contrast with what happens in the classical case. Here we prove
that this is the case as far as $\beta_0<3$. Previously this was
known for integer values of $\beta_0\leq 4$ or for real values
$\beta_0<2$. We also expect our result to be the best one achievable
with variants of the present technique.

We now describe more in detail our assumptions and compare the present
result with the previous ones.

Precisely we assume that $\forall k\geq0$, the following estimates are
fulfilled
\begin{align}
\label{a0}
\left|\partial_x^k  W_0(x,\omega t)\right|&\sleq \langle
x\rangle^{\beta_0-k}\ ,\quad \beta_0< 2\ell-1 \\ ,
\\
\label{a1}
\left|\partial_x^k W_1(x,\omega t)\right|&\sleq \langle
x\rangle^{\beta_1-k}\ ,\quad \beta_1\left\{\begin{matrix}
\leq \ell & {\rm if} & \ell \in [1,2)
  \\
<2(  \ell-1) & {\rm if} & 2\leq \ell 
\end{matrix}\right.\ .
\end{align}
The main point is that the functions $W_i$ are here allowed to grow in
a much faster way than in previous papers (see \cite{2} for the best
previous results).

In literature (see e.g. \cite{hero,1,BGMR2}) perturbations belonging
to a more particular class of symbols are often considered
(cf. Definition \ref{Sm.1} below). In this case we get here a result which
is probably optimal.

The problem of reducibility of equations of the form of
\eqref{schro} has a long history, and the main results have been
obtained in \cite{C87,DS96,DSV02,Kuk93,BG01,LY10,1,2} (see \cite{2} for a
more detailed history). We also mention that our result is limited to
the one dimensional case, while some results on this
problems in more then one dimension have been recently obtained
\cite{GP16,BGMR1,Mon17a, FGMP}. We also recall that related techniques have been
used in order to get a control on the growth of Sobolev norms in
\cite{BGMR2,Mon17}.

We now describe the proof of our result. We recall that the results of
\cite{1,2} were obtained by developing the ideas of
\cite{PT01,IPT05,BBM14}, namely by exploiting pseudodifferential
calculus in order to conjugate the Hamiltonian to a new one which is a
smoothing perturbation of a time independent operator and than
applying a KAM-reducibility scheme, which reduces {\it quadratically}
the size of the perturbation, in order to complete the reduction to
constant coefficients of the system. More applications of these ideas
can be found in several papers (see e.g. \cite{FP, MonK, BM16,
  Giu, BBHM}). In the present paper, in order to prove our reducibility
result, we proceed as follows: first, by a Gauge transformation, we
eliminate from the perturbation the terms containing first order
derivatives. Then we develop a variant of the theory of \cite{1,2} in
order to reduce the perturbation to a smoothing one. The main
difference is that here we do not eliminate time from the normal form
that we construct. More precisely, we first use the theory of \cite{2}
(a variant of Theorem 3.19 of that paper) in order to conjugate
\eqref{H} to a system which is a perturbation of $H_0$ belonging to a
better class of symbols (essentially those considered in \cite{1}) and
then we apply the theory of \cite{BGMR2} in order to conjugate the so
obtained system to another one which is a smoothing perturbation of a
diagonal time dependent system. Finally we eliminate time from the
latter system by an explicit transformation which is done at the
quantum level. Actually, we recall that in \cite{1,2} the main
limitation to the order of the perturbation came from the construction
of the transformation eliminating time from the perturbation.

In Section \ref{stat} we give a precise statement of our main result
and Sect. \ref{proof} contains its proof. Sect. \ref{proof} is split
into 4 subsections: in Subsection \ref{sect.symb} we give some
preliminaries, in Subsection \ref{gauge} we eliminate  $W_1$, in
Subsection \ref{sec.smo} we give some smoothing theorems reducing the
system to a time dependent normal form. Finally in Subsection \ref{eli}
we eliminate time from the normal form and conclude the proof.

\noindent{\it Acknowledgment.} Dario Bambusi was supported by
Universit\'a Statale di Milano and by GNFM. Riccardo Montalto was supported by the Swiss National Science Foundation, grant {\it Hamiltonian systems of infinite dimension}, project number: 200020--165537.

\section{Statement of the main result}\label{stat}

Concerning the potential, when $\ell>1$, we assume that
\begin{equation}
\label{symV}
V(x)=V(-x)\ ,
\end{equation}
and that it admits an asymptotic expansion of the form
\begin{equation}
\label{quasi.1}
V(x)\sim|x|^{2\ell}+\sum_{j\geq 1} V_{2(\ell-j)}(x)\ ,
\end{equation}
with $V_a$ homogeneous of degree $a$, namely s.t., $V_a(\rho
x)=\rho^aV(x)$, $\forall \rho>0$. 
\noindent
We also assume that 
\begin{align}
\label{V3}
V'(x)\not=0\ ,\quad \forall x\not=0\ .
\end{align}

In the case $\ell=1$ we assume that $V(x)=x^2$.

We denote by $\lambda_j^v$ the sequence of the eigenvalues of
\begin{equation}
  \label{H0}
H_0:=-\partial_{xx}+V(x)\ ,
\end{equation}
and remark that they form a sequence $\lambda_j^v\sim c j^d$, with
$d=\frac{2\ell}{\ell+1}$. 
In what follows {\it we will identify $L^2$ with $\ell^2$ by introducing
  the basis of the eigenvector of $H_0$. We also define a reference
  operator $K_0:=H_0^{\frac{\ell+1}{2\ell}}$}

\begin{definition}
  \label{sob}
{For $s\geq 0$, we define the spaces $\cH^s:=D(K_0^s)$
  (domain of the $s$- power of the operator
  $K_0$) endowed by the graph norm}. 
For negative $s$, the space $\cH^s$ is the dual of $\cH^{-s}$.

We will denote by $\cB(\cH^{s_1};\cH^{s_2})$ the space of bounded linear
operators from $\cH^{s_1}$ to $\cH^{s_2}$.
\end{definition}

\begin{remark}
  \label{rem.norm}
Given a function $u \in {\cal H}^s$, one has that 
  \begin{equation}\label{modo 1 norma s}
  \| u \|_{{\cal H}^s} \simeq    \| u \|_{H^{\frac{\ell + 1}{ \ell} s}} + \| \langle x \rangle^{ (\ell + 1) s } u \|_{L^2}
  \end{equation}
  where $H^s$ is the standard Sobolev space and $\norm{.}_{H^s}$ the
  corresponding norm.
\end{remark}

We come to the assumptions on the perturbation. To specify them define first of
all the class of symbols

\begin{definition}
\label{SmV}
The space $S^m_V$ is the space of the functions (symbols) $g\in C^\infty(\R)$ such
that $\forall k\geq 0$ there exists $C_{k}$ with the
property that
\begin{equation}
\label{sm}
\left|\partial^{k}_{x}g(x)\right|\leq C_{k}\langle
x\rangle^{m-k}\ ,\quad \langle x\rangle:=\sqrt{1+x^2}\ .
\end{equation}
\end{definition}

The frequencies $\omega$ will be assumed to vary in the set 
$$
\Omega:=[1,2]^{n}\ ,
$$
or in suitable closed subsets
$\widetilde\Omega$. We will denote by $|\widetilde\Omega|$ the measure
of the set $\widetilde\Omega$. 

Our main result is the following Theorem whose proof will occupy the
rest of the paper.

\begin{theorem}
\label{m.1}
Assume that $W_0\in C^\infty(\T^n;S^{\beta_0}_V)$, $W_1\in
C^\infty(\T^n;S^{\beta_1}_V)$ are real valued and define 
 $$
\beta:=\max\left\{\beta_0,[\beta_1+1]\right\}
$$
where $[\beta_1+1]:=\max\{\beta_1+1,0\}$. Assume $\beta<2\ell-1$ and $\beta_1\leq \ell$.

Then there exists $C,\epsilon_*>0$ and $\forall
\left|\epsilon\right|<\epsilon_*$ a closed set
$\Omega(\epsilon)\subset\Omega$ and, $\forall
\omega\in\Omega(\epsilon)$ there exists a unitary (in $L^2$) time
quasiperiodic map $U_\omega(\omega t)$ s.t. defining $\varphi$ by
$U_\omega(\omega t)\varphi=\psi $, it satisfies the equation
\begin{equation}
\label{rido}
\ir \dot \varphi= H_{\infty}\varphi\ ,
\end{equation}  
with $H_{\infty}=\diag (\lambda_j^\infty)$, with
$\lambda_j^\infty=\lambda_j^\infty(\omega,\epsilon)$ independent of time and 
\begin{equation}
\label{dla}
\left|\lambda_j^\infty-\lambda_j^v\right|\leq C\epsilon j^{\frac{
     \beta}{ \ell +1}}\ .
\end{equation}
Furthermore one has
\begin{itemize}
\item[1.]
  $\displaystyle{\lim_{\epsilon\to0}\left|\Omega-\Omega(\epsilon)\right|=0}$; 
\item[2.] $\forall s,r\geq0$, $\exists \epsilon_{s,r}>0$ and $s_r$ s.t., if
  $|\epsilon|<\epsilon_{s,r}$ then the map
  $\phi\mapsto U_\omega(\phi)$ is of class
  $C^r(\T^n;B(\cH^{s+s_r};\cH^{s}))$; when $r=0$ one has $s_0=0$.
\item[3.] 
$\exists b>0$ s.t.  $\norma{U_\omega(\phi)-\uno}_{B(\cH^{s+\beta};\cH^{s})
}\leq C_s\epsilon^b$.
\end{itemize}
\end{theorem}

As usual, boundedness of Sobolev norms and pure
point nature of the Floquet spectrum follow.

\begin{remark}
  \label{ext}
Actually the result holds for perturbations belonging to a more
general class of symbols. See Definition \ref{Sm} below.
\end{remark}

\begin{remark}
  \label{ext.1}
For perturbations $W$ belonging to a more particular
class of symbols (see Definition \ref{Sm.1} below), the same
conclusion holds for perturbations of order (as defined again in
Definition \ref{Sm.1}) $\beta<2\ell$. The result is probably optimal
within such a class of symbols.
\end{remark}

\section{Proof}\label{proof}

First remark that, given a Schr\"odinger equation 
\begin{equation}\label{blabla}
\ir \partial_t \psi = H(\omega t) \psi
\end{equation}
and a quasi periodic family of unitary transformations $U(\omega t)$, under the change of coordinates $\psi = U(\omega t) \varphi$, the system \eqref{blabla} transforms into the system $\ir \dot \varphi = H_+(\omega t) \varphi$ where
\begin{equation}\label{campo trasformato}
H_+(\phi) = U_{\omega*} H(\phi) := U(\phi)^{- 1} \Big( H(\phi) U(\phi) - \ir \omega \cdot \partial_{\phi} U(\phi) \Big)
\end{equation}

\subsection{Symbols}\label{sect.symb}

To start with, we recall a few classes of symbols essentially
coinciding with those introduced in \cite{2} (see also
\cite{hero,1}). Define
\begin{equation}
\label{lam}
\lambda(x,\xi):=\left(1+\xi^2+|x|^{2\ell}\right)^{\frac{1}{2 \ell}}\ .
\end{equation}
\begin{definition}
\label{Sm}
The space $S^{m_1,m_2}$ is the space of the symbols $g\in C^\infty(\R^2)$ such
that $\forall k_1,k_2\geq 0$ there exists $C_{k_1,k_2}$ with the
property that
\begin{equation}
\label{sm1}
\left|\partial^{k_1}_\xi\partial^{k_2}_{x}g(x,\xi)\right|\leq
C_{k_1,k_2} \left[\lambda(x,\xi)\right]^{m_1-  \ell  k_1}\langle x\rangle^{m_2-k_2}\ .
\end{equation}
\end{definition}

First we remark that $S^{m_1,m_2}\subset S^{m_1+[m_2],0}$
and $S^{m}_V\subset S^{0,m}$.

To a symbol $g\in S^{m_1,m_2}$, we associate its Weyl quantization, namely
the operator $Op^w(g)$, defined by
\begin{equation}
  \label{weyl}
  Op^w(g)\psi(x):=\frac{1}{2\pi}\int_{\R^2}e^{\ir(x-y)\cdot
  \xi}g\left(\frac{x+y}{2};\xi\right) \psi(y)dyd\xi\ .
\end{equation} 
\begin{definition}
  \label{opsmm}
An operator $G$ will be said to be pseudodifferential of class
$OPS^{m_1, m_2}$ if there exists a symbol $g\in S^{m_1,m_2}$ such that
$G=Op^w(g)$. 
\end{definition}

\begin{remark}
  \label{prod}
If $W\in S^m_V$ is a function,  by direct computation one has
  \begin{equation}
    \label{xiW}
Op^w(\xi W(x))=-\im
W\partial_x-\frac{\im}{2}W_x=\frac{-\im\partial_x\circ
  W-W\im\partial_x}{2} \ .
  \end{equation}
\end{remark}

In particular $H_0$ is the Weyl quantization of the symbol
\begin{equation}
  \label{hzero}
h_0(x,\xi):=\xi^2+V(x)\ ,
\end{equation}
and $H$ is the Weyl quantization of  
\begin{equation}
\label{h}
h(x,\xi,\omega t)=(\xi-\epsilon W_1(x,\omega t))^2+ V(x)+\epsilon
W_0(x,\omega t) \ .
\end{equation}

It is well known that given two symbols $a\in S^{m_1,m_2}$ and $b\in
S^{m_1',m_2'}$ there exists a symbol $g\in S^{m_1+m_1',m_2+m_2'}$ such
that $Op^w(a)Op^w(b)=Op^w(g)$, One denotes $a\sharp b:=g$,
furthermore, one has $g=ab+ S^{m_1 + m_1' - \ell , m_2 + m_2' - 1}$ and there exists a full
asymptotic expansion of $a\sharp b$. Furthermore the symbol $(a\sharp
b-b\sharp a)/\ir$ of $1/\ir$ times the commutator of the two Weyl
operators is called Moyal Bracket of $a$ and $b$ and will be denoted
by $\left\{a;b\right\}_M$. It turns out that
$S^{m_1+m_1'- \ell ,m_2+m_2'-1}\ni
\left\{a;b\right\}_M$ furthermore, one has $
\left\{a;b\right\}_M
=\left\{a;b\right\}+S^{m_1 + m_2 - 2 \ell, m_1 + m_2 - 2}$, where
$\{.;.\}$ denotes the Poisson Bracket. 

The application of the Calderon Vaillancourt Theorem yields the
following Lemma.

\begin{lemma}
\label{caderon}
Let $g\in S^{m_1,m_2}$, then one has
\begin{equation}
\label{CV}
Op^w(g)\in \cB(\cH^{s_1+s};\cH^{s})\ ,\quad \forall s\ ,\quad
\forall s_1\geq m_1+[m_2]\ .
\end{equation}
\end{lemma}

In the proof we will also need the classes of symbols used in
\cite{1}, thus we recall the corresponding definitions

\begin{definition}
\label{Sm.1}
The space $S^{m}$ is the space of the symbols $g\in C^\infty(\R^2)$ such
that $\forall k_1,k_2\geq 0$ there exists $C_{k_1,k_2}$ with the
property that
\begin{equation}
\label{sm.1}
\left|\partial^{k_1}_\xi\partial^{k_2}_{x}g(x,\xi)\right|\leq
C_{k_1,k_2} \left[\lambda(x,\xi)\right]^{m-  \ell k_1-  k_2}\ .
\end{equation}
\end{definition}
Given a symbol $g \in S^m$, we say that the corresponding Weyl operator ${\rm Op}^w(g)$ belongs to the class $OPS^m$.

In order to deal with functions $p$ which depend on $(x,\xi)$ through
$h_0$ only, namely such that there exist a $\tilde p$ with the
property that
$$
p(x,\xi)=\tilde p(h_0(x,\xi))\ ,
$$
we introduce the following class of symbols. 

\begin{definition}
\label{d.sm}
A function $\tilde p\in C^{\infty}$ will be said to be of class
$\widetilde S^m$
if one has
\begin{equation}
\label{sm.12}
\left|\frac{\partial^k\tilde p}{\partial E^k}(E)\right|\sleq \langle
E\rangle^{\frac{m}{2\ell}-k} \ .
\end{equation} 
\end{definition}

Sometimes symbols of this class are also called classical symbols. 

By abuse of notation, we will
say that $p\in\widetilde S^m$ if there exists $\tilde p\in \widetilde
S^m$ s.t. $p(x,\xi)=\tilde p(h_0(x,\xi))$. We say that the corresponding Weyl operator ${\rm Op}^w(p)$ belongs to the class $\widetilde{OPS}^m$.


\subsection{Reduction of $W_1$}\label{gauge}

\begin{lemma}
\label{mag}
There exists $b\in C^{\infty}(\T^n;S^{[\beta_1+1]}_V)$ s.t. the transformation
\begin{equation}\label{mag.1.1}
\begin{aligned}
& \Umag (\phi) : \psi(x) \mapsto e^{-\ir \epsilon  b(\phi , x)} \psi(x)  
\end{aligned}
\end{equation}
conjugates \eqref{H} to
\begin{equation}
\label{H1.1}
H^{(1)}(\phi ):=- \partial_{xx}+V(x)+ \epsilon W_0^{(1)}(\phi, x)\,,
\quad W_0^{(1)} := W_0-\omega \cdot \partial_\phi b \in
C^\infty(\T^n, S_V^{ \beta})
\end{equation}
where $ \beta := {\rm max}\{  \beta_0,[ \beta_1 +
  1] \} < 2 \ell - 1$.

If $\beta_1\leq \ell$ then \eqref{mag.1.1} maps $\cH^s$ into itself.
\end{lemma}

Remark that $W^{(1)}_0\in S^{0,\beta}$.

\proof 
One has
\begin{equation}\label{regole coniugazione gauge}
\begin{aligned}
& [\Umag] ^{- 1} \circ( \ir  \partial_x) \circ \Umag  & =  \ir \partial_x
-   \epsilon  b_x  \,, 
\\
& - [\Umag]^{- 1} \circ \Big(  \ir \omega \cdot \partial_\phi   \Umag \Big)  & =
-  \epsilon   \omega \cdot \partial_\phi b\,,
\end{aligned}
\end{equation}
while the operators of multiplication are invariant under under the
transformation (with $\phi$ considered as a parameter). Thus, if we
define $b$ by
\begin{equation}
\label{defb}
b(\phi, x )=  \int_0^x W_1(\phi, y)dy\ ,
\end{equation} 
we get $b\in S^{[\beta_1+1]}$ and $ \ir\partial_x+ W_1(x,\phi)$ is
conjugated to the differential operator $ \ir\partial_x$ and thus
\eqref{H} is conjugated to \eqref{H1.1}.

In order to show that the spaces $\cH^s$ are left invariant by the
transformation generated by $b$ we apply Theorem 1.2 of \cite{MaRo}
according to which it is enough to verify that $[b,K_0]K_0^{-1}$ is a
bounded operator. This is easily verified by remarking that, its
principal symbol is
$$
\{b,h_0^{\frac{\ell+1}{2\ell}}\} h_0^{-\frac{\ell+1}{2\ell}}\ ,
$$
and thus, by explicit computation of the Poisson bracket, one gets
that this the symbol of a bounded operator provided
$\beta_1\leq \ell$.
 \qed

\subsection{Smoothing theorems}\label{sec.smo}

The conjugation of $H^{(1)}$ to a Hamiltonian with a smoothing
perturbation is obtained through
the combination of a few smoothing theorems which essentially have
already been proved in previous papers, but are here combined in a new
way. For the proof we mostly refer to the original papers.

{\it In the case $\ell=1$ Theorem \ref{m.1} follows from Theorem 2.4 of
  \cite{2}, so {\bf we concentrate on the case $\ell>1$}}.

The first result that we need is a smoothing theorem which is a variant of
Theorem 3.19 of \cite{2}.

\begin{theorem}
\label{smoothing}
[Smoothing Theorem 1] Consider the Hamiltonian \eqref{H1.1} and assume
$$
\beta<2\ell-1\ ;
$$
fix an arbitrary
$\kappa>0$, 
then there exists a time dependent family of unitary transformations
$\uquat(\phi)$ which transform the Hamiltonian \eqref{H1.1} into
a pseudo-differential operator $H^{(2)}$ with symbol
$h^{(2)}$ given by
\begin{equation}
\label{hreg}
h^{(2)}(\phi, x,\xi)=h_0(x,\xi)+\epsilon z^{(2)}(h_0(x,\xi),\phi)+\epsilon r^{(2)} (h_0(x,\xi),\phi)
\end{equation}
where $z^{(2)}\in C^{\infty}(\T^n;\tilde S^{\beta})$ is a function of
$(x,\xi)$ through $h_0$ only, while the remainder fulfills 
\begin{equation}
  \label{rsto.1}
r^{(2)}\in C^{\infty}(\T^n;S^{-\kappa,0})\ .
\end{equation}
Furthermore, one has
\begin{itemize}
\item[1.] $\forall r\geq0$, $\exists s_r$ s.t., the map
  $\phi\mapsto \uquat(\phi)$ is of class
  $C^r(\T^n;B(\cH^{s+s_r};\cH^{s}))$; when $r=0$ one has $s_0=0$.
\item[2.]
$\norma{\uquat(\phi)-\uno}_{B(\cH^{s+\beta};\cH^{s})
}\leq C_s\epsilon$.
\end{itemize}
\end{theorem}

The proof is essentially identical to the proof of Theorem 3.19 of
\cite{2}, the difference is that one makes the first transformation
reducing \eqref{H1.1} to the form (3.41) of \cite{2} and then, instead
of eliminating the time dependence from the average of $W^{(1)}_0$,
one iterates the previous step (as in \cite{BGMR2}), thus getting a
normal form which is a function of time, but depending on the space
variables through $h_0$ only. The main point is that this normal form
constitute a new time dependent perturbation {\it which is of
  class $C^\infty(\T^n;\tilde S^\beta)$}.

\begin{remark}
\label{limitation}
The limitation $\beta<2\ell-1$ is needed in the proof of the above
theorem. In particular it is needed in order to ensure that, if
$\chi\in S^{\beta-\ell+1,0}$ and $\Phi^t_{\chi}$ is the corresponding
Hamiltonian flow, then, given a symbol $f$ of some class,
$f\circ\Phi^t_{\chi}$ is also a symbol of the same class. 
\end{remark}

We now apply Theorem 3.8 of \cite{BGMR2} which gives. 

\begin{theorem}
\label{thm:smooth1}[Smoothing Theorem 2]  Under the assumptions of Theorem
\ref{smoothing}.\\ There exists a unitary (time-dependent) operator
$U^{(3)}(\phi)$ in $L^2(\R)$ which transforms $H^{(2)}$ (and thus the
Hamiltonian \eqref{H}) into the Hamiltonian
\begin{equation}
\label{hN}
H^{(3)}(\phi) := H_0+\epsilon
Z^{(3)}(\phi) +\epsilon R^{(3)} (\phi)
\end{equation}
where $Z^{(3)}(\phi) \in
C^\infty(\T^n, \widetilde{\ops}^{\beta} )$  commutes with $H_0$, i.e. $[
  Z^{(3)}(\omega t) , H_0 ] = 0$, while   $R^{(3)}\in C^\infty(\T^n,
\ops^{-\kappa,0})$.
Furthermore, one has
\begin{itemize}
\item[1.] $\forall r\geq0$, $\exists s_r$ s.t., the map
  $\phi\mapsto U^{(3)}(\phi)$ is of class
  $C^r(\T^n;B(\cH^{s+s_r};\cH^{s}))$; when $r=0$ one has $s_0=0$.
\item[2.]
$\norma{U^{(3)}(\phi )-\uno}_{B(\cH^{s+\beta};\cH^{s})
}\leq C_s\epsilon$.
\end{itemize}
\end{theorem}

\proof First we recall that according to Theorem (7-8) of \cite{hero}
there exists a pseudodifferential operator $Q\in\ops^{-(\ell+1)}$ s.t.
\begin{equation}
  \label{k_0}
H_0=K_1^{\frac{2\ell}{\ell+1}}+Q\ ,\ \text{and}\ [K_1;Q]=0\ ,
\end{equation}
and the spectrum of $K_1$ is $\{j+\sigma\}_{j\geq 0}$ with\footnote{The difference between the operator $K_1$
  and the operator $K_0$ introduced previously, is that the spectrum
  of $K_0$ is asymptotically equal to $j+\sigma$, while the spectrum
  of $K_1$ is exactly equal to $j+\sigma$}
$\sigma>0$. Remark that $K_1$ and $Q$ are diagonal on the basis of the
eigenfunctions of $H_0$. Then Theorem 3.8 of \cite{BGMR2} applies and
gives the result with $Z^{(3)}$ which commutes with $K_1$, however,
since the eigenvalues of $K_1$ are simple, $Z^{(3)}$ commutes also
with $H_0$. \qed

\begin{remark}
  \label{autoval}
By the previous theorem, the matrix of the operator $Z^{(3)}(\omega
t)$ is diagonal on the basis of the eigenfunctions of $H_0$. Thus on
this basis
\begin{equation}
  \label{muj}
Z^{(3)}(\phi)=diag(\mu_j(\phi))
\end{equation}
with suitable smooth functions $\mu_j(\phi)$ which satisfy for any $m \in \N$ the estimate $|\mu_j|_{C^m(\T^n)} \leq C_m j^{\frac{\beta}{\ell + 1}}$ for a suitable constant $C_m > 0$. 
\end{remark}

We are now going to show that, due to the property that
$Z^{(3)}(\omega t)$ is a pseudodifferential operator, the
$\mu_j\null's$ are essentially smooth functions of the
eigenvalues of $H_0$, i.e. of $\lambda_j^v$.

\begin{lemma}
  \label{interpolation}
For any $\kappa$ there exists a smooth function $\langle
z^{(3)}\rangle\in C^{\infty}(\T^n;\tilde S^{\beta})$ and a sequence of
functions $\delta_j(\phi)$ s.t.
\begin{equation}
  \label{autoval1}
\mu_j(\phi)= \langle z^{(3)}\rangle(\lambda_j^v,\phi)+\delta_j(\phi)\ ,
\end{equation}
and, for any $m\geq 0$, there exist $C_m$ s.t.
\begin{equation}
  \label{est}
\left\vert\delta_j\right\vert_{C^m(\T^{n})}\leq C_m j^{-\kappa}\ .
\end{equation}
\end{lemma}
\proof Denote by $z^{(3)}$ be the symbol of $Z^{(3)}$ (where
we drop the dependence on $t$). Let $\eta(E)$ be a smooth compactly
supported function and write
$$
z^{(3)}=z^{(3)}_0+z^{(3)}_R\ ,
$$
where $z^{(3)}_R(x,\xi,\omega t):=z^{(3)}(x,\xi,\omega
t)\eta(h_0(x,\xi)) $ and $z^{(3)}_0:=z^{(3)}-z^{(3)}_R$.

By the commutation property one has
\begin{equation}
  \label{moyal}
\left\{ z^{(3)};h_0 \right\}_M=0\ \Longrightarrow \left\{
  z^{(3)}_0;h_0 \right\}=:\delta \in S^{\beta-\ell-3} \ .
\end{equation}
Denote by $\Phi^t_{h_0}$ the flow of the Hamiltonian system with
Hamiltonian $h_0$ and define the average of $z^{(3)}_0$ by
\begin{equation}
  \label{media}
\langle z^{(3)}_0 \rangle (x,\xi,\phi):=\left.\frac{1}{T(E)}\int_0^{T(E)} z^{(3)}_0(
\Phi^\tau_{h_0} (x,\xi),\phi)d\tau\right|_{E=h_0(x,\xi)} \ , 
\end{equation}
where $T(E)$ is the period of the classical orbits of $h_0$ at energy
$E$. By Lemma 4.16 of \cite{1}, one has $\langle z^{(3)}\rangle\in
C^{\infty}(\T^n, \tilde S^\beta)$. 

Define now 
$$
\check z^{(3)}_0:= z_0^{(3)} -\langle z^{(3)}_0\rangle\ ,
$$
Remark that the average of $\delta$ vanishes. So that $\check z^{(3)}_0$
is the only solution with zero average of the equation
$$
\left\{ h_0; \check z^{(3)}\right\}=\delta\ .
$$ Now, according to Lemma 4.17 of \cite{1}, such solution is of class
$S^{\beta-2(\ell+1)}$. It follows that $\check z^{(3)}_0\in
S^{\beta-2(\ell+1)} $.  Furthermore, by standard argument one has that
$ \langle z^{(3)}_0 \rangle (x,\xi)=\langle
z_0^{(3)}\rangle(h_0(x,\xi),\phi) $ depends on $(x,\xi)$ through
$h_0$ only. Finally, by functional calculus one has that the Weyl operator
of $\langle z^{(3)}_0\rangle(h_0,\phi)$ is given by
\begin{equation}
  \label{ultim}
\langle z^{(3)}_0\rangle( H_0)+\ops^{\beta-(\ell+1)}\ .
\end{equation}
Thus one has
$$
Z^{(3)}=\langle z^{(3)}_0\rangle( H_0)+\ops^{\beta-(\ell+1)}\ .
$$ Repeating the argument for $Z^{(3)}-\langle z^{(3)}_0\rangle(
H_0)$, and iterating, one gets the result. \qed

\subsection{Elimination of time from $Z^{(3)}$ and preparation for KAM theory}
\label{eli}

In this section we eliminate time from $Z^{(3)}$ and we get a system
suitable for the application of the KAM Theorem 7.3 of \cite{1}.

First we fix a $\tau>n-1$ and define the set $\Omega_\gamma$ of
Diophantine frequencies with constant $\gamma$ by
\begin{definition}
  \label{dioph}
  The frequencies $\omega$ belonging to the set
  \begin{equation}
    \label{dio}
\Omega_\gamma:=\left\{ \omega\in[1,2]^n \ :\ \left|\omega\cdot
k\right|\geq\frac{\gamma}{\left| k\right|^\tau}\ ,\ \forall k\in
\Z^n\setminus\{0\} \right\}
  \end{equation}
  are called Diophantine.
\end{definition}

It is well known that $\left|\Omega-\Omega_\gamma\right|\leq C
\gamma$ for a suitable positive constant $C$. 

In the following we will denote by
$Lip\left(\Omega_\gamma;C^r(\T^n;\cB(\cH^s;\cH^{s'}))\right)$ the
space of Lipschitz functions from $\Omega_\gamma$ to
$C^r(\T^n;\cB(\cH^s;\cH^{s'}))$.  

\begin{lemma}
  \label{time}
Define
$$
\bar z^{(3)}(E):=\frac{1}{(2\pi)^n}\int_{\T^n}\langle
z^{(3)}\rangle(E,\phi)d^n\phi \ ,
$$
then $\bar z^{(3)}\in \widetilde S^\beta$ and, for $\omega\in
\Omega_\gamma$, there exists a unitary (time-dependent) operator
$U^{(4)}(\omega t)$ in $\cH$ which transforms \eqref{hN} into
\begin{equation}
  \label{hsm}
H^{(4)}(\phi):=A_0+\epsilon R_0(\phi)\ ,
\end{equation}
where 
\begin{align}
\label{hN1}
A_0 := diag(\lambda^{(0)}_j(\omega))\ ,
\\
\label{lzeroj}
\lambda^{(0)}_j(\omega)=\lambda_j^v+\bar z^{(3)}(\lambda_j^v)\ .
\end{align}
Furthermore, one has
\begin{itemize}
\item[1.] $\forall r\geq0$, the map
  $\phi\mapsto U^{(4)} (\phi)$ is of class
  $C^r(\T^n;B(\cH^{s};\cH^{s-\beta r}))$. 
\item[2.]
$\norma{U^{(4)}(\phi)-\uno}_{\cB(\cH^{s+\beta};\cH^{s})
}\leq C_s\epsilon$. 
\item[3.] For all $r$ one has
  $R_0:=Lip\left(\Omega_\gamma;C^r(\T^n;\cB(\cH^s;\cH^{s+\kappa-\beta 
  r-1}))\right) $.  
\end{itemize}
\end{lemma}
\proof The transformation is obtained by eliminating time from
$H_0+\epsilon Z^{(3)}$. To this end remark that, since $Z^{(3)}$ is
diagonal, the Schr\"odinger equation with such Hamiltonian is a
sequence of scalar equations of the form
$$
\ir\dot\psi_j=\lambda_j^v\psi_j+\epsilon \mu_j(\omega t)\psi_j\ ,
$$ where, according to \eqref{autoval1}, $\mu_j$ has a smooth
dependence on $\lambda_j^v$ according \eqref{autoval1}.

In order to eliminate time dependence from this equation consider the
Fourier expansion
$$
\mu_j(\phi)=\sum_{k\in\Z^n}\mu_{j,k}e^{\ir k\cdot\phi}\ ,
$$
and define
$$ c_j(\phi)=\sum_{k\not=0}\frac{\mu_{j,k}}{\ir\omega\cdot k}e^{\ir
  k\cdot\phi}\ .
$$
Then the transformation $\utre$ is defined by $\psi_j\mapsto
e^{-\epsilon \ir
  c_j(\omega t)}\psi_j$. A simple analysis of the transformation
shows that it fulfills the properties 1-3.\qed 

\vskip10pt

As a final step we state a corollary which shows that $H^{(4)}$
fulfills the assumptions of Theorem 7.3 of \cite{1} which thus gives
the result.
\begin{corollary}
\label{Hop}
For any positive $\gamma$, $r$, $\kappa$, there exists a set
$\Omega^{(0)}_\gamma\subset \Omega_\gamma$, positive $a$ and
$\epsilon_*$ s.t., if $|\epsilon|< \epsilon_*$ then, for any
$\omega\in\Omega^{(0)}_\gamma$, the unitary (in $L^2$) operator
$U_1:=U^{(1)}\circ U^{(2)}\circ U^{(3)}\circ U^{(4)} $
conjugates \eqref{H} to \eqref{hsm}, furthermore, the following
properties hold 
\begin{align}
\left|\Omega_\gamma\setminus \Omega_\gamma^{(0)}\right|\leq
C\gamma^a\ ,
\\
\label{A}
A_0:=\diag(\lambda_j^{(0)})\ ,
\end{align}
with $\lambda_j^{(0)}=\lambda_j^{(0)}(\omega)$ Lipschitz dependent on
$\omega\in\Omega_\gamma$ and fulfilling the following inequalities
(with $d=2\ell/(\ell+1)$)
\begin{align}
\label{diaga.1}
\left|\lambda_j^{(0)}-\lambda_j^v\right|\sleq j^{\frac{
    \beta}{l+1}}\ ,
\\
\label{diaga.2}
\left|\lambda_i^{(0)}-\lambda_j^{(0)}\right|\sgeq
\left|i^d-j^d\right|\ ,
\\
\label{diaga.3}
\left|\frac{\Delta(\lambda_i^{(0)}-\lambda_j^{(0)})}{\Delta\omega}
\right| 
\sleq
\epsilon|i^d-j^d| \ .
\\
\label{diaga.303}
\left|\lambda_i^{(0)}-\lambda_j^{(0)}+\omega\cdot k\right|\geq
\frac{\gamma\sidjd}{1+|k|^\tau}\ ,\quad
\left|i-j\right|+\left|k\right|\not=0\ ,
\end{align}
where, as usual, for any Lipschitz function $f$ we denoted $\Delta
f=f(\omega)-f(\omega')$.

Furthermore, one has
\begin{itemize}
\item[1.] the map
  $\phi\mapsto U_1(\phi)$ is of class
  $C^r(\T^n;\cB(\cH^{s};\cH^{s-\beta r}))$. 
\item[2.]
$\norma{U_1 (\phi)-\uno}_{\cB(\cH^{s+\beta};\cH^{s})
}\leq C_s\epsilon$. 
\item[3.] One has
  $R_0\in Lip\left(\Omega_\gamma;C^r(\T^n;\cB(\cH^s;\cH^{s+\kappa}))\right) $.  
\end{itemize}
\end{corollary}

We end the section by remarking that Theorem \ref{m.1} is now an
immediate consequence of the Theorem 7.3 of \cite{1}. 

\addcontentsline{toc}{chapter}{Bibliography}

\bibliography{biblio}
\bibliographystyle{alpha}
\def\cprime{$'$}


\end{document}